\documentclass[ preprint]{aastex}
\usepackage{epsfig,rotate}
\usepackage{graphicx}
\newcommand{\be}{\begin{equation}}
\newcommand{\ee}{\end{equation}}

\newcommand{\icar}{{\it Icarus, }}

\def\be{\begin{equation}}
\def\ee{\end{equation}}
\def\beq{\begin{equation}}
\def\eeq{\end{equation}}

\def\bea{\begin{eqnarray}}
\def\eea{\end{eqnarray}}
\def\noi{\noindent}

\begin{document}

\title
{Disk Planet Interactions and Early Evolution in Young Planetary Systems}

\author{J.C.B. Papaloizou}
\affil{Astronomy Unit,\\  Queen Mary, University of London,\\
Mile End Rd.,\\  London, E14NS.}

\email{jcbp@maths.qmul.ac.uk}



\begin{abstract} 
We study and review  disk protoplanet interactions using local 
shearing box simulations. These suffer
the disadvantage of having potential artefacts arising
from  periodic boundary conditions but the advantage, when compared
to global simulations, of being
able to capture much of the dynamics close to the protoplanet
at high resolution for low computational cost.
 Cases with and without self
sustained  MHD turbulence
are considered.
The conditions for gap formation and the transition 
from type I migration  are investigated and found to 
depend on whether the single parameter 
$M_p R^3/(M_* H^3),$ with $M_p, M_*, R,$ and $H$ being the protoplanet mass,
the central mass, the orbital radius and the disk semi-thickness respectively
exceeds a number of order unity.
We also investigate the coorbital torques experienced by a moving protoplanet
in an inviscid disk.
This is done by demonstrating the equivalence of the problem for a moving
protoplanet to one where the protoplanet is in a fixed orbit which the
disk material flows through radially as a result of the action of
an appropriate external torque.
For sustainable
coorbital torques to be realized a quasi steady state must be realized
in which the planet migrates through the disk without accreting
significant mass. In that case, although there is 
sensitivity to computational parameters, in agreement with earlier
work by Masset \& Papaloizou (2003) based on global simulations,
the coorbital torques are
proportional to the migration speed and result in a positive
feedback on the migration, enhancing it and potentially leading to a runaway.
This could lead to fast migration for protoplanets in the Saturn mass
range in massive disks and may be relevant to the mass period correlation
for extrasolar planets which gives a preponderance of sub Jovian masses
at short orbital periods.  

\end{abstract}



\section{Introduction}\label{S0} 
The ongoing discovery of extrasolar  planets has resulted
in increased investigation of theories of planet formation
(e.g. Mayor \& Queloz 1995;
Marcy, Cochran, \& Mayor 2000; Vogt et al. 2002).
Possible formation scenarios for gas giant planets  are either through
direct gravitational instability occurring in
a young protostellar disk (e.g. Boss 2001) or through 
the accumulation of a solid  core that
undergoes   rapid gas accretion once   its mass has 
reached a critical value
$\simeq 15$ Earth masses (e.g. 
Pollack et al. 1996). In either case planetary formation is  likely
to  be initiated at significantly greater orbital radii than those
currently  observed implying that a process of orbital
migration has brought  them closer to the central star.
Disk protoplanet interaction provides a natural  migration mechanism.

A protoplanet exerts torques on a protostellar disk
through the excitation of spiral density waves 
(e.g. Goldreich \& Tremaine 1979;
Papaloizou \& Lin 1984; Lin \& Papaloizou 1993).
These waves carry   away either positive or negative angular momentum 
which is deposited in the disk  at the locations
where the
waves are damped. As a result of this, a negative torque is exerted
on the
protoplanet by the outer disk and a positive torque 
is exerted on it by the disk interior to its orbit.

In recent years much numerical
work on disk protoplanet interactions
has been undertaken for protoplanets with a range
of masses in both laminar and turbulent disks
in which the turbulence is maintained by the magnetorotational
instability
(eg. Bryden et al. 1999; Kley 1999; Lubow, Seibert, \&
Artymowicz 1999,
D'Angelo, G., Henning, \& Kley, W., 2002, Nelson \& Papaloizou 2003,
Winters Balbus \& Hawley 2003a, Papaloizou Nelson \& Snellgrove 
2004; Nelson \& Papaloizou 2004).
This work indicates that a
sufficiently massive protoplanet can open up an annular gap in the disk
centred on its orbital radius.
For typical protostellar disk models gap formation at $5AU$  starts to
occur for protoplanet masses of around a Saturn mass with the gap
becoming deep for a Jovian mass.

For low mass disks gap formation, or the transition
from linear to non linear disk response, is associated with the transition
between type I migration (eg. Ward 1997) and type II migration
(eg. Lin \& Papaloizou 1986). For both of these regimes the
time scale of the migration in standard disk models,
with mass comparable to that of the minimum mass solar nebula,
is shorter than the disk lifetime making it a threat to the
survival of protoplanets (see eg. Terquem Papaloizou \& Nelson 2000
for a review) as well as a mechanism for producing close
orbiting giant planets. 

More recently Masset \& Papaloizou (2003) considered  a form of
potentially very fast migration induced by the action
of coorbital torques acting on disk material as it passes
through the coorbital region of a migrating protoplanet.
This is in contrast to torques produced by waves dissipating
in the disk away from the orbit. They found using two dimensional
global simulations at rather low resolution that such torques
could lead to a positive feedback and a fast migration for
typically Saturn mass protoplanets in massive disks.
This may provide a mechanism for bringing sub Jovian 
mass objects close to the star which, if they are then
slowed down and undergo
some additional accretion may become hot Jupiters.  

In this paper we explore some calculations of disk
planet interactions and related migrational torque calculations using local 
shearing box simulations. These focus on a local patch on the disk
and so may be done at high resolution for relatively low cost.
We consider both non turbulent disks and disks with turbulence driven
by the magnetorotational instability or MRI (Balbus \& Hawley 1991).
This is the most likely mechanism producing angular
momentum transport and thus  the '{\it viscous}'   evolution of the disk
that ultimately drives type II migration (Lin \& Papaloizou 1986).

We discuss the noisy type I migration in turbulent
disks and the condition for gap formation.
We also explore at length coorbital torques associated with migrating
protoplanets and find that the local box simulations are  consistent
 with the lower resolution global ones.  

The plan of the paper is as follows.
In section \ref{S1}
we give the basic equations and describe the
local shearing box model in frames that may be either uniformly rotating 
or migrating with an angular velocity which is a function of time. 
In section \ref{S3} we 
describe the numerical procedure and models simulated.
In section \ref{S4} we go on to
describe results of simulations with protoplanets and  the calculation
of the
forces acting 
on them that lead to migration. In particular we examine the coorbital
torque generated by material as it passes through the coorbital region
from one side of the protoplanet to the other and present numerical
results for two different flow through speeds.
Finally in section \ref{S5} we discuss our results
and their implications for extrasolar planetary configurations.

\section{Basic Equations and  Model set up}\label{S1} 

In this paper we describe local simulations
of protoplanets interacting with a Keplerian disk flow
with and without 
MHD turbulence driven by the MRI. We consider simulations 
for which the magnetic field, when present has zero net flux so that 
an internally generated dynamo is maintained. Simulations with turbulence
require high resolution with usually not less than $32$ grid cells per scale
height $H.$ They are greatly facilitated by considering
a local shearing box 
rather than a complete disk (Goldreich \& Lynden-Bell 1965). 

\noindent
The governing equations for  ideal MHD written in an  inertial frame 
are:
\begin{equation}
\frac{\partial \rho}{\partial t}+ \nabla \cdot {\rho\bf v}=0, \label{cont}
\end{equation}
\begin{equation}
\rho \left(\frac{\partial {\bf v}}{\partial t}
 + {\bf v}\cdot\nabla{\bf v}\right)= 
-\nabla p -\rho \nabla\Phi +
\frac{1}{4\pi}(\nabla \times {\bf B}) \times {\bf B}, \label{mot}
\end{equation}
\begin{equation}
\frac{\partial {\bf B}}{\partial t}=\nabla \times ({\bf v} \times {\bf B}).
\label{induct}
\end{equation}
where ${\bf v}, P, \rho, {\bf B}$ and $\Phi$ denote the fluid
velocity, pressure, density  and  magnetic field respectively.
The gravitational potential is $\Phi.$ It contains contributions
from a central mass $M_*$ and an orbiting secondary mass or planet
$M_p.$
Thus adopting a cylindrical coordinate system $(r, \varphi, z)$
centred
on $M_*,$
\be \Phi=-GM_*/r 
 -{GM_p\over \sqrt{r^2 + R^2 -2r R \cos(\varphi-\varphi_s)+ z^2 +b^2}},\ee
 where $(R, \varphi_s , 0)$ are the coordinates of the secondary,
$M_p$ is
the   secondary  mass and $G$  the gravitational constant.

\subsection{The Local Shearing Box}
Following Goldreich \& Lynden-Bell (1965)
we consider a small Cartesian box centred on a point at which the Keplerian
angular velocity is $\Omega_p.$
Normally one takes the centre of the box to be at a fixed point
in the underlying disk flow which corresponds to
the motion of a secondary mass in circular orbit. However, it is possible
 to consider a box centred on a point with changing radius $R(t)$
and varying angular velocity $\Omega_p(t) = \sqrt{GM_*/R(t)^{3}},$ 
$M_*$ being the central mass and $G$ being the gravitational constant.
For a box centred on the secondary mass this corresponds to the situation
when the secondary migrates.
We also suppose that as the point moves the  
disk semi-thickness is $H(t)$ and 
that this determines the radial scale of the flow.
 
We use the local Cartesian coordinates
$x = r- R(t),$ $z,$ and the mutually orthogonal coordinate
$y,$ to define local dimensionless 
Cartesian coordinates $(x' , y' , z')=
( x/H(t), y/H(t), z/H(t))$ with associated
unit vectors $({\bf i}, {\bf j}, {\bf k}).$ 

\noindent The direction ${\bf i}$
is outwards along the line joining the origin
to the central object while ${\bf k}$ points in the vertical direction.
The direction ${\bf j}$ is that of the unperturbed Keplerian shear flow.
In this non inertial frame the flow velocity is ${\bf u}.$
It is also convenient to use a rescaled time
$t'$ such that $d t' = \Omega_p(t)dt.$

 The idea behind using the local box is that there is a small
parameter $ h = H/R,$ measuring the aspect ratio of the disk.
We adopt a spatially isothermal equation of state $P =\rho c_s^2,$
with $c_s$ being the sound speed which may depend on time, such that
the aspect ratio is also equal to the
ratio of sound speed to orbital velocity so that
$h = c_s/(R\Omega_p).$

 For a moving centre we suppose that there is another
small parameter $|\epsilon| = |dR/dt/(\Omega_p R)| \sim  |dH/dt/(\Omega_p H)|.$
This measures the ratio of the orbital time to
the time for the centre to migrate significantly in the radial direction.

\noindent We shall assume that $|\epsilon| /h = O(h^{k}),$
  for some $ 1 > k  > 0.$ This enables us to
 retain terms of order $\epsilon /h = dR/dt/(\Omega_p H)$
but neglect terms of order $\epsilon$  and higher.

\subsection{Dimensionless Variables}
 We introduce a dimensionless velocity, sound
speed and planet mass through
${\bf u}' = {\bf u}/(\Omega_p H),$ $c_s' = c_s/(\Omega_p H),$ and
$q = M_p R^3/(M_* H^3)$ respectively.
As we do not include  the self-gravity of the disk in calculating
its response, the magnitude of the disk density may be arbitrarily
scaled. All the simulations presented here  start with a uniform density
$\rho_0$ which, together with $H,$ may be used to specify the mass scale.
We define a dimensionless  density and magnetic field through
$\rho' = \rho /\rho_0$ and
${\bf B}' =  {\bf B}/(\Omega_p H \sqrt{4\pi\rho_0}).$
The equation of motion may be written in terms of these
dimensionless variables, after taking the limit
$\epsilon \rightarrow 0,$ and $h \rightarrow 0$
simultaineously but retaining the ratio $\epsilon /h$ in the form

\begin{equation}
\frac{\partial {\bf u}'}{\partial t'}
+{\bf u}'\cdot\nabla{\bf u}'
+ 2{\bf {\hat k}}{\bf \times}{\bf u}' -3x'{\bf{\hat i}}= -{\epsilon{\bf{\hat j}}
\over 2h}
-{\nabla (\rho' c_s^{'2}) \over \rho'} -\nabla\Phi_p' +
{(\nabla \times {\bf B}') \times {\bf B}' \over \rho'}, \label{boxmotdim}
\end{equation}
where
\be \Phi_p' = -{q \over \sqrt{x^{'2} + y^{'2}  +  z'^{2} + b^{'2}}},
\label{dimpot}\ee
with dimensionless softening parameter  $b' = b/H$
and of course the spatial derivatives are with respect to the dimensionless 
 coordinates.
Similarly the dimensionless induction equation becomes
\begin{equation}
\frac{\partial {\bf B}'}{\partial t'}=
\nabla \times ({\bf u}' \times {\bf B}').
\label{inductdim}
\end{equation}
and the dimensionless continuity equation is
\begin{equation}
\frac{\partial \rho'}{\partial t' }+ \nabla (\cdot \rho' {\bf u}') = 0.
\label{contdim}
\end{equation}

We  note that the term $\propto x'$ in equation (\ref{boxmotdim})
is derived from a first order Taylor expansion  about $x=0$
of the combination
of the
gravitational  acceleration due to the central mass and the centrifugal
acceleration (Goldreich \& Lynden-Bell 1965).
However, in order to reduce computational requirements,
in the work presented here, we have neglected the $z'$ dependence
of the gravitational potential due to the secondary
 given by equation (\ref{dimpot})
and also  of that due to the central mass.
Therefore vertical
stratification is neglected.
Simulations  of unstratified
boxes are often undertaken 
in MHD because they contain the essential
physics and ease computational requirements
(eg. Hawley, Gammie \& Balbus 1995).

\noindent In the steady state  box with a fixed centre and
no planet or magnetic field, the equilibrium
velocity is due to Keplerian shear
and thus
\be {\bf u'} = (0,-3 x'/2 ,0). \label{boxv}\ee

\subsection{Steady and Moving Frames}
When the centre is fixed so that $\Omega_p,$ $R$ and $H$ are constant,
equations (\ref{boxmotdim}- \ref{contdim}) 
are the standard equations for a shearing box
(Goldreich \$ Lynden-Bell 1965).

When the centre migrates with $\Omega_p,$ $R$ and $H$
being functions of time, in the limit of small, $\epsilon$ and $h$
but $\epsilon /h$ retained, the equations are the same but for one added
dimensionless acceleration $ -\epsilon {\hat j}/(2h).$
This corresponds to the dimensionless torque required to
make the disk gas move with speed $-dR/dt$ in the absence
of the secondary. This situation means that a migrating
planet moving through non migrating gas with speed $dR/dt$ 
is equivalent 
to a non migrating planet immersed in a disk with gas given a torque
which results in it migrating with speed $-dR/dt.$ 
This symmetry has already been exploited without justification  in two dimensional global
simulations by Masset \& Papaloizou (2003).
However, the approximation scheme requires $\epsilon$ and $h$ to be small.
This means that the migration
time to move through $R$ measured by
$1/ \epsilon$  has to be very  long compared to the orbital time.
But the time to migrate through $H,$ which is measured by
 $h/ \epsilon $  can be significantly faster.
 
\subsection{Boundary Conditions}

The shearing box is presumed to represent a local patch of a differentially
rotating disk. Thus the appropriate boundary conditions on the bounding  faces
$y = {\rm constant} = \pm Y$ and $z = {\rm constant} = \pm  Z$ come
from the requirement of periodicity in the local Cartesian coordinate
directions normal to the boundaries.
On the boundary faces $x = {\rm constant} = \pm X,$ 
the boundary  requirement is for periodicity in local shearing coordinates.
Thus for any state variable
$F(x,y,z),$ the condition is that $F(X,y,z) = F(-X, y -3Xt' ,z).$
This means that information on one radial boundary face is communicated
to the other boundary face at a location in the azimuthal coordinate
$y$  shifted by  the distance
the faces have sheared apart since the start of the simulation
(see Hawley, Gammie \& Balbus 1995).

In dimensionless coordinates
the boundaries of the box are $x' =\pm X/H,$  $y' =\pm Y/H,$ $z' =\pm Z/H.$
Also if the magnetic field has zero net flux and is dynamo generated and
thus a spontaneous product of the simulation, given that
 the only parameter occurring in equations 
(\ref{boxmotdim}-\ref{contdim} )
 is the dimensionless mass $q = M_p R^3/(M_* H^3),$
we should be able to  consider the dependence of the
time averaged outcome of a simulation to be only on
$ X/H, Y/H, Z/H,$
and $q = M_p R^3/(M_* H^3).$

 \begin{table}
 \begin{tabular}{ccccc}
 \hline
       &       &        &                 &          \\
 Model & $ q $ &  $b/H$ & $|\epsilon|/h$ & ${\bf B}$ \\
       &       &        &                 &            \\
 \hline
  A    & $0.0$ &   - & $0.0$ & NO \\
  B    & $0.1$ & $0.3$ & $0.0$ & NO \\
  C    & $0.1$ & $0.3$ & $0.0$ &YES \\
  D    & $0.3$ & $0.3$ & $0.0$ &YES \\
  E    & $1.0$ & $0.3$ & $0.0$ &YES \\
  F    & $2.0$ & $0.3$ & $0.0$ & NO \\
  G    & $2.0$ & $0.3$ & $3/(8\pi)$ & NO \\
  H    & $2.0$ & $0.3$ & $3/(64\pi)$ & NO \\
  I    & $2.0$ & $0.6$ & $3/(8\pi)$ & NO \\
  J    & $2.0$ & $0.6$ & $3/(64\pi)$ & NO \\
 \hline
 \hline
\end{tabular}
\caption{ \label{table1}
The first column gives the model label, the second the value of $q,$
the third gives the softening parameter $ b/H,$ the fourth
gives $|\epsilon|/h$ the magnitude of which gives the ratio of the induced
disk flow through speed to the sound speed and
the fifth column indicates whether a magnetic field was present.
Model A was a reference model used to check that a box with
no protoplanet and no magnetic field did not show any evolution
or instabilities.}
\end{table}

Note that the above discussion implies that for a box of fixed dimension, the
only distinguishing parameter is $ q =  M_p R^3/(M_* H^3).$ 
This parameter may also be interpreted as the cube of the  ratio of the Hill
radius to disk scale height and the condition that  $q > \sim  1$ 
leads to the thermal condition of 
Lin \& Papaloizou (1993) that the Hill radius should exceed the disk
scale height for gap formation to occur.  
 From Korycansky \& Papaloizou (1996)
this condition is also required in order that the perturbation
due to the protoplanet be non linear.

\begin{figure}
\centerline{
\epsfig{file= 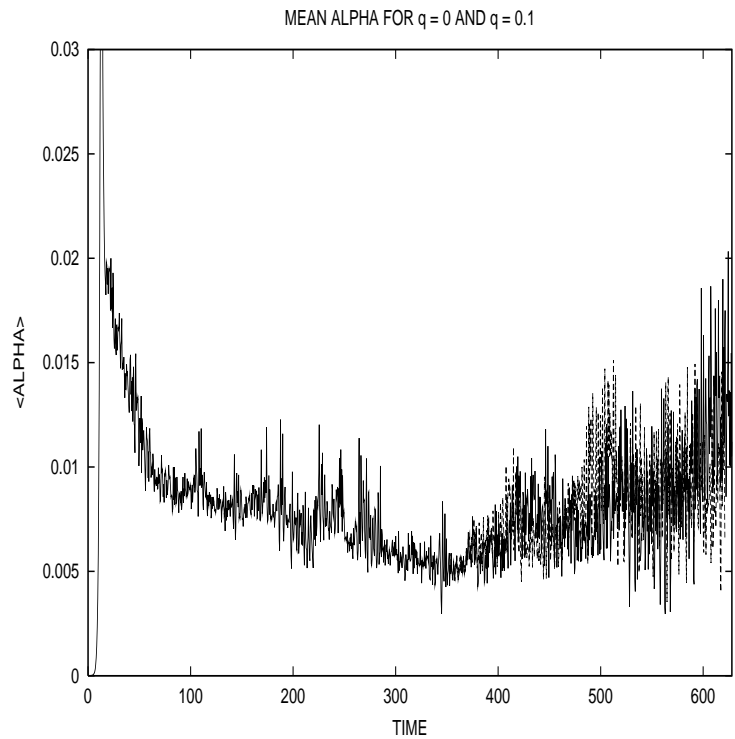 , height=10.cm,width=10.cm,angle =270}}
\caption[]{The   volume averaged stress parameter $\langle \alpha \rangle,$
is plotted as a function of a dimensionless
time for no planet, $q=0$ at all times
and for the case
when the planet corresponding to $q=0.1$  was inserted at time $353.$
Although both curves are independently  very noisy after this time
they cannot be separated. }
\label{fig1}
\end{figure}

\begin{figure}
\centerline{
\epsfig{file= 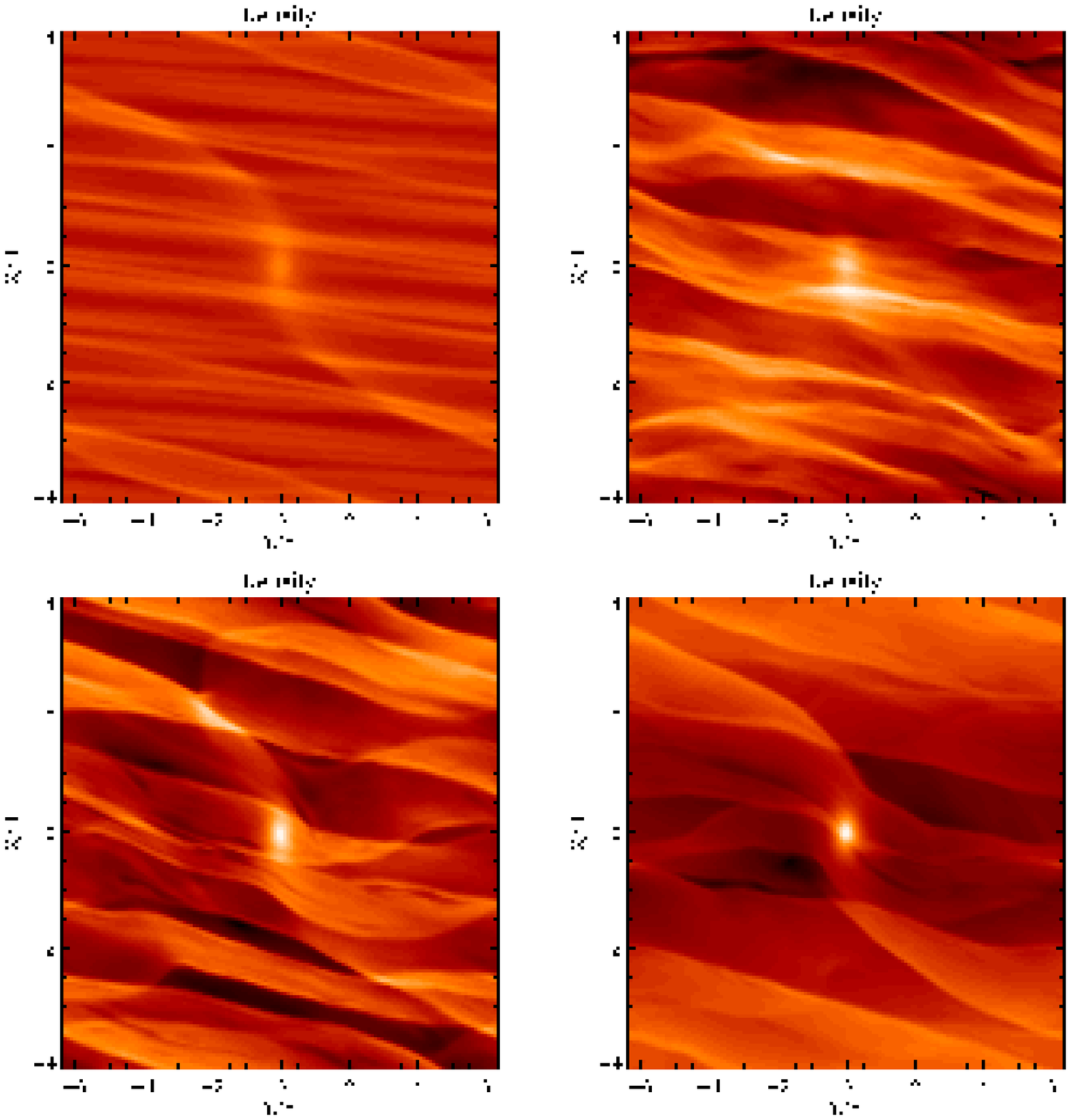 , height=10.cm,width=10.cm,angle =0}}
\caption[]{Typical mid plane
 density contour plots  near the end of the simulations
for upper left panel, $q =0.1$ and no magnetic field (model B), upper right panel, $q=0.1$  with magnetic
field (model C), lower left panel, $q =0.3$ with magnetic field (model D) and 
lower right panel, $q=1.0$ with magnetic field (model E).
 The ratio of maximum to minimum densities in these
plots were 1.42, 2,9, 3.27, and  11.22 respectively. In the case of the upper left panel,
the ratio was reduced to 1.25
 if the wakes produced by the protoplanet were excluded.}
\label{fig23}
\end{figure}

\section{Numerical Procedure and Models Simulated} \label{S3}
\noindent
The numerical
method  is that of  characteristics
constrained transport MOCCT  (eg. Hawley \& Stone 1995).
The code used has been developed from  a version
of NIRVANA originally written by U. Ziegler
(see Ziegler \& R\"udiger 2000 and
references therein). It has been used in a number of simulations of 
disk planet interactions in two and three dimensions with and
without magnetic fields
(eg. Kley 1999;  Nelson et al. 2000; Steinacker \& Papaloizou 2002;
 Papaloizou Nelson \& Snellgrove  2004).

All of the simulations were for shearing boxes with $X=4H, Y = 2\pi H,$
and $ Z= H/2.$ These boxes are larger than the standard one often used in
MHD simulations (eg. Hawley Gammie \& Balbus 1995)
which has $X = H/2, Y = \pi H$
and $ Z= H/2.$ Such a larger box is required to cover the scale
of the disk planet interaction ( Papaloizou Nelson \& Snellgrove 2004;
Nelson \& Papaloizou 2004). As in those works, the simulations
were carried out with $261,$ $200,$  and $35$
grid points in the $x',$ $ y' $ and $z'$ directions respectively.
At this resolution, MHD turbulence and associated dynamo can be maintained
for beyond one hundred orbits of the box centre.
Some parameters associated with the simulations are given in table I

In order to investigate conditions for gap formation we have run models
with a variety of values of $q =M_p R^3/(M_* H^3)$ with and without magnetic fields.
We have also studied the torques acting on a  planet moving outwards by applying a torque
to the disk material to make it flow inwards (see discussion above).
Values of $|\epsilon|/h$ are given in table I.
But note that because of the symmetry of the box, equivalent results are obtained
if the flow is in the opposite direction. 

The softening parameter should be chosen so as
to represent the effects of vertical stratification and the expected value is expected to
be $\sim H$ (eg. Papaloizou 2002; Masset 2001;  Nelson \& Papaloizou 2004).
The value adopted here was usually $b=0.3H.$

At this point a cautionary note should be inserted. For sustainable
coorbital torques to be realized a quasi steady state must be realized
in which the planet migrates through the disk with a mass consistent
with that assumed when calculating the tidal interaction.
If a very small softening parameter is used, the diverging point
mass potential coupled with a fixed low temperature
isothermal equation of state  results in arbitrarily
large amounts of mass being deposited on the planet.
Apart from preventing states with a steady flow through the coorbital region
being attained, large amounts of accretion onto the planet
cannot be handled correctly in calculations such as those 
 performed  here and elsewhere for disk planet interactions
that neglect self-gravity of the disk and the proper
physics of  protoplanetary structure.
Steady coorbital torques require a planet  that self-consistently
 does not increase
its mass significantly through disk mass accretion as can be modeled
with a softening parameter $b \sim H$ such as adopted here.

But note that there is some expected sensitivity of torques  to the  
value  of the softening
parameter (eg. Artymowicz 1993;  Papaloizou 2002; Nelson \& Papaloizou 2004).
Accordingly for purposes of comparison some models were run with $b=0.6H.$ 
We found negligible mass accumulation on the protoplanet for $b=0.6H$
but a more noticeable effect when $b=0.3H.$

Because they are not dimensionless and  can be scaled away the following
numerical values  have no significance, but  for convenience we adopted 
$h = 0.01,$ $\rho_0 = 0.001,$ $GM_*=1$ and $R =1.$
All our results   below are obtained   using these values.

\noindent The gravitational potential was flattened ( made to attain a constant value
in a continuous manner)
at distances exceeding $3H$ from the protoplanet in order to satisfy the 
periodic boundary conditions. Tests with larger flattening distance
have indicated indicated insensitivity to the precise choice.

\begin{figure}
\centerline{
\epsfig{file=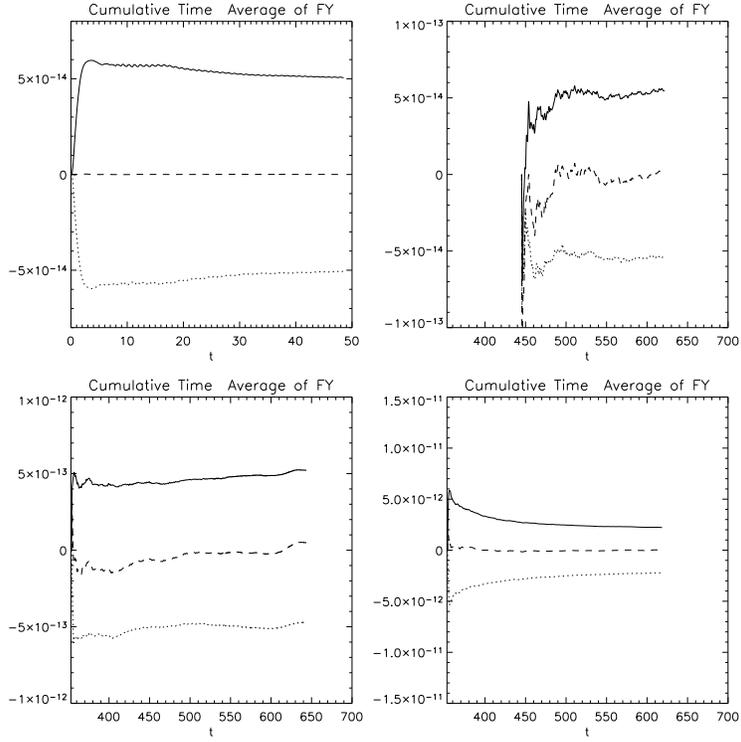, height=10.0cm,width=10.cm,angle =0} }
\caption[]{ Running time averages of the force acting 
on the planet in the direction
of orbital motion plotted as a function of dimensionless time. These are for, 
 the upper left panel, $q =0.1$ and no magnetic field (model B), upper right panel, $q=0.1$  with magnetic
field (model C), lower left panel, $q =0.3$ with magnetic field (model D) and
 lower right panel, $q=1.0$ with magnetic field (model E).
For each panel the upper curve represents the force due to material
exterior to the protoplanet, the lower curve 
 represents the force due to material
interior to the protoplanet while the central curve gives
the total contribution.}
\label{fig9}
\end{figure}

\begin{figure}
\centerline{
\epsfig{file= 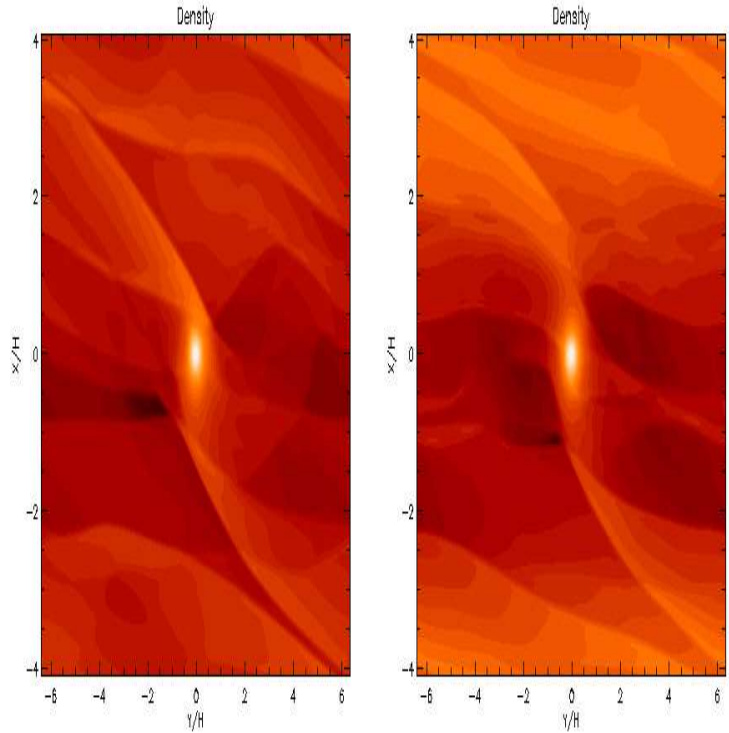 ,height=10.cm,width=10.cm,angle =0} }
\caption[]{Typical mid plane  density contour plots
for left panel $q=2.0$ with torque induced fast flow through (model G)
and right panel $q=2.0$
with a torque induced slower flow through of disk material (model H).
The softening parameter was $b=0.3H.$
In these cases the magnitude of the torque applied to the disk material
was chosen to produce  expected inflow speeds of
$(3/(8\pi))c_s$ and $(3/(64\pi))c_s$ respectively.
These inflow speeds correspond to $|\epsilon|/h = 3/(8\pi)$ (fast flow through)
and $|\epsilon|/h = 3/(64\pi)$ (slower flow through).
  These models had no magnetic field.}
\label{fig01}
\end{figure}

\noindent
\begin{figure}
\centerline{
\epsfig{file=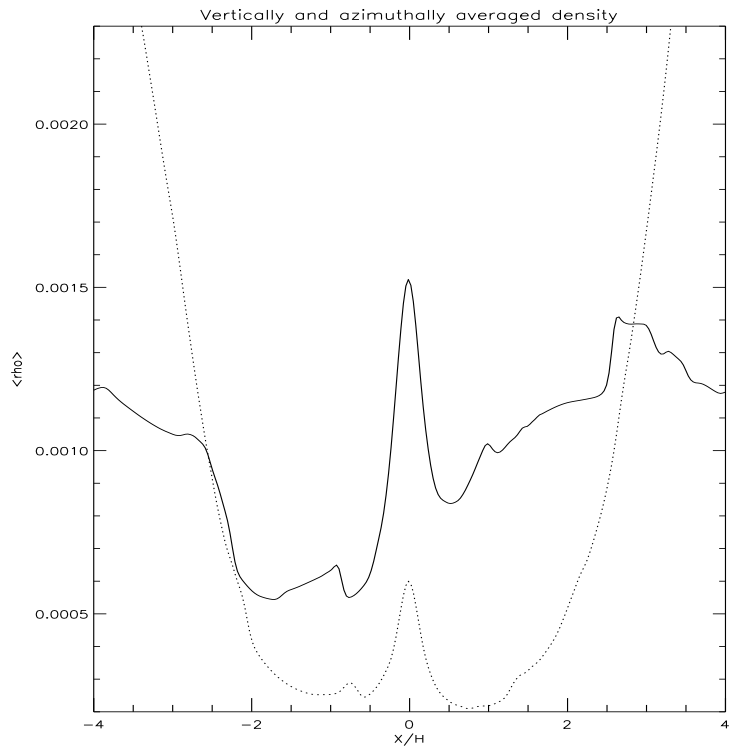,height=10.cm,width=10.cm,angle =0} }
\caption[]{The mean surface density averaged over $y$ and $z$
as a function of radial
coordinate $x/H$ for $q=2.0$ with no flow through (dotted curve, model F)
and with flow through (full curve, model G).
 These models had  $b=0.3H$ and  have no
magnetic field.
}
\label{fig20}
\end{figure}

\noindent
\begin{figure}
\centerline{
\epsfig{file=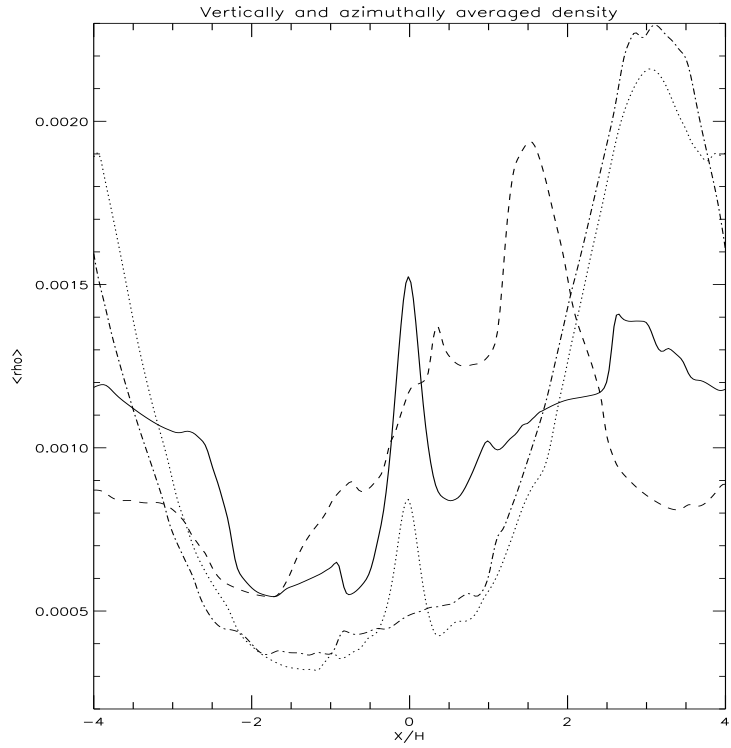,height=10.cm,width=10.cm,angle =0} }
\caption[]{The mean surface density averaged over $y$ and $z$
as a function of radial
coordinate $x/H$ for $q=2.0$ with  fast flow through and $b=0.3H$ 
(model G, full curve),
with slower flow through  and $b=0.3H$ ( model H, dotted curve),
with  fast flow through  and $b=0.6H$ (model I, dashed curve) ,
with slower flow through  and $b=0.6H$ ( model J, dot-dashed curve).
These models have no
magnetic field.
}
\label{fig2}
\end{figure}

Models with magnetic fields all had conserved zero net flux. 
For these the initial field was taken to be vertical, independent of $y$ and $z$ 
and to  vary  sinusoidally in $x$ with a wavelength of $H.$
The amplitude was  chosen to make the  initial ratio 
of the total magnetic energy to volume integrated pressure 
to be $0.0025.$ The initial velocity in the $x$ direction at each
grid point  was
chosen to be the product of  a random number between $-1.0$ and $1.0$
and $0.1c_s.$
Without a perturbing secondary, these models reach a
turbulent state in which the  ratio of volume mean magnetic energy to volume mean pressure
is in the range $0.002 - 0.02$ and the radially or volume averaged   Shakura \& Sunyaev $\alpha$ parameter
is in the range $0.003 - 0.02.$ The time averaged value of the volume averaged $\alpha$ is typically
$0.008$ (see eg. Winters Balbus \& Hawley 2003b, Papaloizou Nelson \& Snellgrove 2004
for details). A plot of the evolution of the volume averaged $\alpha$ for a simulation with no
secondary mass which begins with the above initial
conditions for the magnetic field  is presented in figure \ref{fig1}.
Models with magnetic fields and protoplanets were
initiated by inserting the protoplanet into this
model with established MHD  turbulence  after $353$ time units.

Models without magnetic fields all had perturbing protoplanets
so that these could be initiated from rest. 

\subsubsection{Vertical and Horizontal Averages
}\label{S3a}
\noindent
We find it useful to consider quantities that are
vertically and azimuthally averaged over the $(y ,z)$ domain ${\cal D}.$ 
Sometimes an additional running time average may be adopted.
The vertical and  azimuthal average of some quantity $Q$ is defined  by

\noindent
\begin{figure}
\centerline{
\epsfig{file= 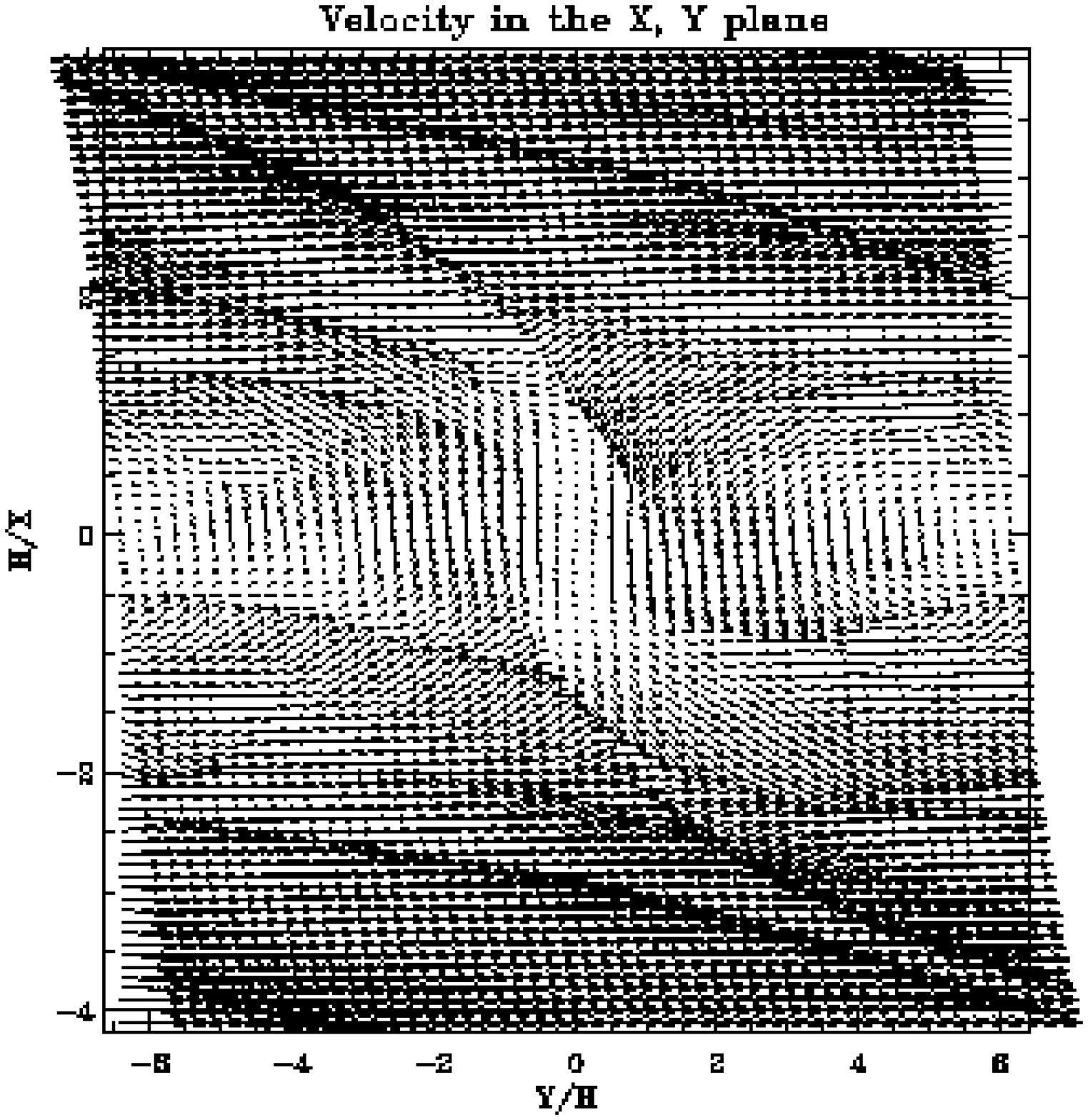 ,height=10.cm,width=10.cm,angle =0} }
\caption[]{ Velocity vectors in the mid plane coorbital region for $q=2.0$
 no magnetic field and no flow through (model F) }
\label{fig10}
 \end{figure}

\begin{equation}
{\overline {Q}} ={\int_{\cal D} \rho  Q dz dy \over \int_{\cal D}  \rho dz dy}.
\end{equation}

\noindent The  disk surface density is given by
\begin{equation}
\Sigma(x,t) = {1\over 2Y }\int_{\cal D} \rho dz dy.
\end{equation}

\noindent The vertically and azimuthally
averaged  Maxwell and
Reynolds stresses,  are  given by:
\begin{equation}
T_M= 2Y
\Sigma{\overline{\left({B_x B_y \over 4\pi\rho}\right)}}
\end{equation}
and
\begin{equation}
T_{Re}=2Y
\Sigma
{\overline{\delta v_x\delta v_y}}.
\end{equation}
 Here the velocity fluctuations $\delta v_x$ and $\delta v_y$
are defined through,
\begin{equation}
\delta v_x=v_x-{\overline{v_x}},
\end{equation}
\begin{equation}
\delta v_y=v_y- {\overline{v_y}}.
\end{equation}
The horizontally and vertically averaged  Shakura  \& Sunyaev (1973)
$\alpha$ stress parameter appropriate to the
total stress is  given by
\begin{equation}
\alpha=\frac{T_{Re}-T_M}{2Y
\Sigma{\overline{ \left(P/\rho\right)}}},
\end{equation}
The angular momentum flow across a line of constant $x$
is given by

\begin{equation}
 {\cal F} =2Y R \Sigma \alpha\overline{P/\rho}.
\label{transport}
\end{equation}

\begin{figure}
\centerline{
\epsfig{file=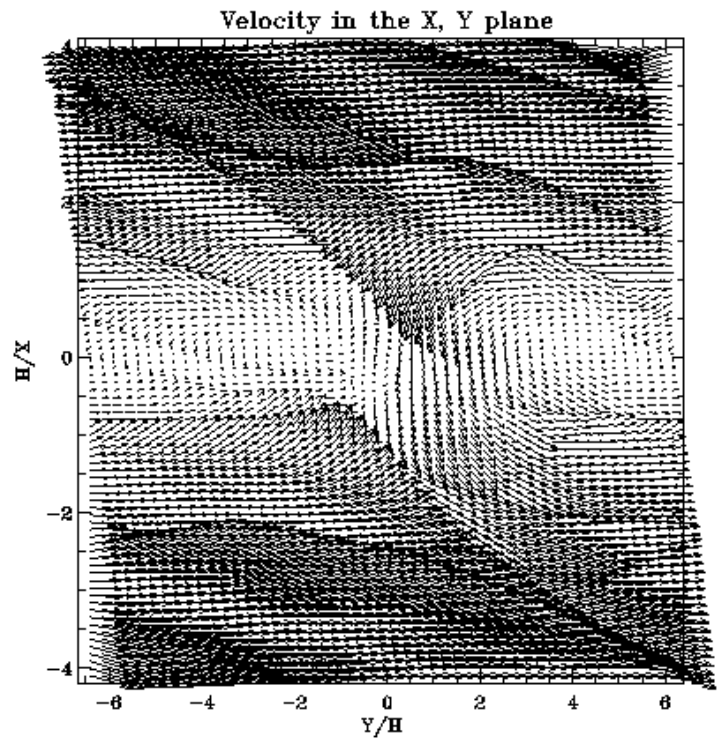,height=10.cm,width=10.cm,angle =0} }
\caption[]
{ Velocity vectors in the  mid plane
coorbital region for $q=2.0,$  $b=0.3H,$
 with  no magnetic field
and fast  flow through by  induced torqued  disk material (model G).
}
\label{fig3a}
\end{figure}

\begin{figure}
\centerline{
\epsfig{file= 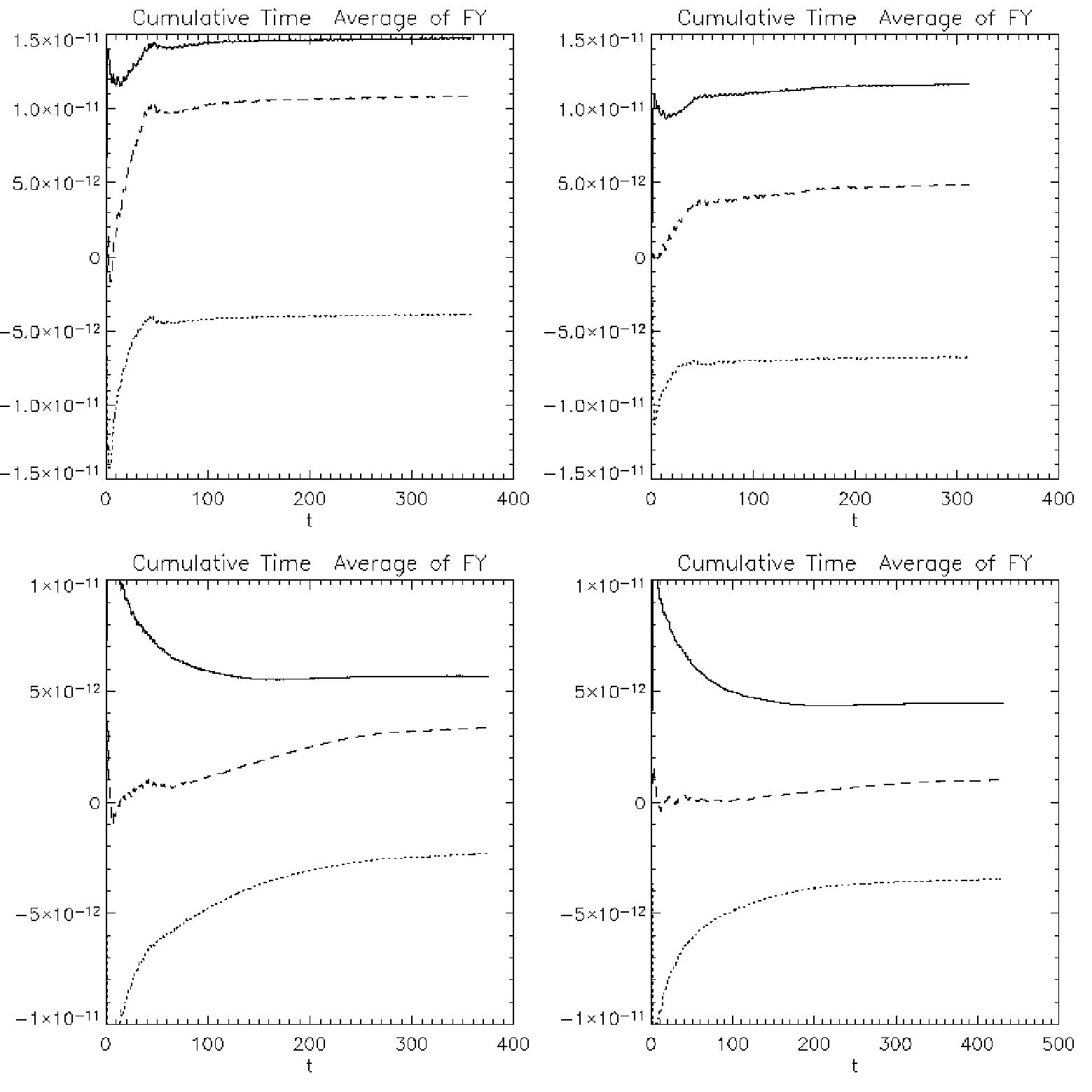, height=10.cm,width=10.cm,angle =0} }
\caption[]
{Running time averages of the force acting on the planet in the direction
of orbital motion as a function of dimensionless time. 
These are for $q = 2.0$   with
fast flow through and $b=0.3H$ ( model G, upper left panel), $q = 2.0$
with fast flow through  and $b=0.6H$ ( model I, upper right panel), $q = 2.0$
with slower flow through  and $b=0.3H$
(model H, lower left panel) , and $q = 2.0$
with slower flow through  and $b=0.6H$ ( model J, lower right panel).
These models have no
magnetic field. For each panel the upper curve
represents the force due to material
exterior to the protoplanet, the lower curve
represents the force due to material
interior to the protoplanet while the central curve gives
the total contribution.}
\label{fig3}
\end{figure}

\section{  Results of Simulations with Protoplanets} \label{S4}
We have performed simulations with fixed protoplanets with
$q=0.1, 0.3, 1.0$ and $2.0.$ with  and without magnetic fields.
As $q$ increases, a gap is formed once $q$ exceeds  a number around unity.
If magnetic fields are present this is also the value beyond which perturbations
from the protoplanet dominate the turbulence.

In figure \ref{fig1} we plot the volume averaged
stress parameter $\langle \alpha \rangle,$
as a function of  dimensionless
time for no planet, $q=0,$ 
together with the same quantity for the run in which a protoplanet with
$q=0.1$  was inserted at time $353.$
Although both curves are independently very noisy after this time
they cannot be separated verifying the fact that a protoplanet with $q=0.1$
is not strong enough to perturb the turbulence significantly. 

Nonetheless all the simulations show the prominent  density wake
associated with it.
Typical  mid plane density contour plots  near the end of the simulations
for upper left panel, model B (q=0.1), upper right panel,  model C (q=0.1), 
lower left panel, model D (q=0.3)  and
lower right panel, model E (q=1) are given in in figure \ref{fig23}.
The ratio of maximum to minimum densities  was 
1.42, 2,9, 3.27, and  11.22 respectively.
When the dominant wakes
produced by the protoplanet are excluded
in model B, this ratio is reduced to 1.25. This  ratio
measures the effect of  waves excited by 
ghost protoplanets in neighbouring boxes
which are there on account of the periodic
boundary conditions in shearing coordinates.
 It indicates that the fluctuations
due to ghost neighbours  are relatively small compared to the effects
due to either magnetic fields or dominant
perturbations close to the protoplanet due ro its own gravity. 

The simulations with $ q < 1$ all have embedded protoplanets
while for $q \ge 1$ a prominent gap is formed (see also  Papaloizou Nelson 
\& Snellgrove 2004). This occurs with and without a magnetic field.
The condition for gap formation that $q >1$ is equivalent to requiring
that The size of the Hill sphere exceed the disk thickness, $H.$ 
The additional 
requirement discussed by Lin \& Papaloizou (1993) is  that tidal forces
should be enough to overcome viscous diffusion is expressed
by $q > 40 \alpha (R/H).$ To compare with the box simulations
considered here, we should set $ 2R = L_y /\pi = 4H$ and
the value of $\alpha$ has to be related to the MHD turbulence
when present. We adopt the value
$\alpha \sim 0.005$  corresponding to a mean value of  the Maxwell
stress. Thus we see that for the MHD cases 
the viscous criterion for gap formation is  satisfied 
for the boxes  when $q > 0.4.$ Thus $q \ge 1$ gives the  expected condition
for gap formation in all the box simulations. 
An extrapolation to make the azimuthal extent of the box equal
to $2\pi R$ would give $q > 0.2(R/H).$
Thus it should be borne in mind that the tendency for the gap to
fill in a full circle may be inderestimated for thin disks  with MHD
 in a local box
of the type we have used.

\subsubsection{Torques and Migration}
An important aspect of the disk protoplanet interaction is that
it produces torques on the protoplanet that can lead to orbital migration.
For a box simulation the torque is $RF_{y}$ where $F_{y}$ is the force
in the $y$ direction. We consider the contributions from the regions
exterior to and interior to the protoplanet separately.
One can see from  eg. figure \ref{fig23} that the density enhancement
in the wake exterior to the protoplanet tends to drag it backwards
while the density enhancement
interior to the protoplanet tends to accelerate it. By  the
 symmetry of the box  these
contributions cancel on average. However, features which break the symmetry
such as a radial flow through can result in a non zero net torque/force.
 
Running time averages of the forces acting
on the planet in the direction
of orbital motion are plotted as a function of dimensionless time in figure
\ref{fig9}. These are for,
the upper left panel, model B ,upper right panel, model C, 
lower left panel, model D, 
lower right panel, model E.
We normalize these using the fiducial form
\be F_{y0} = q^2 \Omega_p^2 H^3 \Sigma. \ee
For our units, this gives
\be F_{y0} = 10^{-11}q^2 . \ee
For $q=0.1$ and $q=0.3$ a gap does not form
and a reasonable fit to the one sided outward force  
produced by regions exterior to the secondary is 
\be F_y = 0.5 F_{y0}.\ee

We remark that the time averaged forces from exterior and interior region
are   very similar with and without turbulence due to a magnetic field.
Because of the symmetry of the shearing box, 
contributions from the regions exterior to the
protoplanet orbit must eventually 
balance the contributions from the interior regions.
However, this may take a long time to achieve in the MHD case 
where noise can produce a ten percent bias for embedded protoplanets
even after fifty orbits.  Effects not yet modelled
such as long timescale turbulent fluctuations in global
disks where there is bias introduced both by the geometry, initial conditions
and external boundary conditions
 may have significant consequences for the net migration of embedded
objects (see Nelson \& Papaloizou 2004).

To compare with the expected rates of type I migration
in a disk without a magnetic field,
(eg. Ward 1997) we assume there is an asymmetry between outward and inward
forces introduced by global geometrical effects of magnitude $C_1 h,$
where $C_1$ is a constant between $1$ and $10$ (Ward 2000).
This is  taken to be in favour of a net  drag force
and  vanishes in the limit $h \rightarrow 0$ as required.
The net force is then
\be F_y = - 0.5 F_{y0}C_1 h.\ee
This leads to an expected inflow time
\be \tau_{mig} = - {R \over dR/dt} =
 \frac{M_*}{M_p} \frac{M_{*}}{{C_1\Sigma  R^2}}
       h^2 \Omega_p^{-1}. \label{boxmig}\ee 

A physical explanation for why the net force amounts to a drag
	can be based on the fact that the wakes originate
as a density wave launched starting from the location 
where the disk shears past the protoplanet with the sound speed.
In a constant aspect ratio disk geometrical effects (not present in the box) 
make this location closer to the protoplanet in the outer
regions resulting in a larger contribution to the net force.
The resulting asymmetry is of order $h.$

The migration time of a non gap forming protoplanet embedded in a three
dimensional disk has been calculated using a linear
analysis by Tanaka,Takeuchi,\& Ward (2002). They derive the following
expression for the migration time of a protoplanet embedded in a
locally isothermal disk
with a  uniform surface density profile:

\begin{equation}
\tau_{mig}= \frac{M_*}{2.7M_p}\frac{M_{*}}{{\Sigma R^2}}
h^2 \Omega_p^{-1} \label{ward-tmig}
\end{equation}
These authors comment
that in general this expression gives a migration time that is a factor
of between 2 -- 3 times slower than similar expressions derived for
flat, two--dimensional disks (e.g. Ward 1997). Interestingly
both  (\ref{boxmig}) and (\ref{ward-tmig})
 agree for $C_1 = 2.7$ 
and correspond to a  rather fast 
inward migration time of $8 \times 10^5 y$ at $5 AU$ for
 $1 M_{\oplus}$ in a gas disk with surface density
chosen so that there is two Jupiter masses within $5 AU.$
But note that the  magnitude is sensitive to the softening parameter
which needs to be comparable to the vertical thickness or $\sim H.$

\subsection{Coorbital Torques}

By coorbital torque we mean a torque exerted
on the protoplanet 
by disk material flowing through the orbit. This has been discussed recently
by Masset \& Papaloizou (2003).
We have shown that in the local approximation adopted here
 the torque on a  slowly radially migrating protoplanet
is the same as that due to disk material
radially migrating with the same speed  in the opposite direction 
flowing past a protoplanet in fixed orbit.

As mentioned above, in the absence of migration the symmetry
of the shearing box results in zero net torque on the protoplanet.
However, introducing a migration direction breaks the symmetry
and results in a non zero torque on the protoplanet
which changes sign with the direction of migration.
From very general considerations we expect the net torque
to be proportional to the migration velocity.
This can result in either positive feedback when
the torque assists the protoplanet migration or negative feedback
when it acts in reverse.

Masset \& Papaloizou (2003) argue for positive feedback because for an outwardly
migrating protoplanet material traverses the coorbital zone
from outside to inside. As it does so it moves along the outer boundary
of  the coorbital zone occupied by material librating around the coorbital 
equilibrium (Lagrange points) passing close to the protoplanet.
As this passage occurs angular momentum is transferred to the protoplanet,
a process which acts to assist the migration of the protoplanet
and which accordingly gives a positive feedback.
We now look at this process in more detail.

The expected force exerted on the secondary by material flowing through the coorbital region
is estimated as follows. The mass flow through the box is at rate
$ 2Y\Sigma dR/dt.$ The outward momentum per 
unit mass imparted to the secondary when
the disk matter moves across the coorbital 
region is $w\Omega_p/2,$ where $w$ is it's
width. For this we adopt $w=2H$ (see below ).
Then the rate of transfer of outward momentum to the planet is
$F_{cr} = 2Y\Omega_p H \Sigma (dR/dt).$ This  can be expressed as

\be F_{cr} = 2Y\Omega_p^2 H^2 \Sigma ((dR/dt)/c_s) .\ee    
This may also be written in the form

\be F_{cr} =  {1\over 2} M_d \Omega_p  (dR/dt) , \label{crt00}\ee
where $ M_d =  4Y \Sigma H$
is the disk mass that would fill the coorbital zone of width
$2H$  were it to do so at the background surface density.

However, torques on the protoplanet do not only arise from material
passing through the coorbital zone. Material that is forced to
comove with the protoplanet, either because it has been accreted by
it, or because it librates about the coorbital equilibrium
points has to be supplied with the same force per 
unit mass as the protoplanet, by the protoplanet
 so as to maintain its migration.
This results in an additional force acting on the protoplanet given by

\be F_{crb} =  -{1\over 2} M_b \Omega_p (dR/dt) ,\ee
where $ M_b$ is the coorbital bound mass.
The total force so far  is thus

\be F_{cr} =  {1\over 2}( M_d - M_b) \Omega_p (dR/dt) . \label{crto}\ee

However, there are other forces. Because of the breaking
of the symmetry of the box, the forces acting due to density
waves from the two sides do not necessarily cancel.
In this context note that these is an asymmetry in the surface density profile.
Thus we should expect a further wave torque component
which is proportional to the migration speed.
The indication from our simulations is that this acts
as a drag on the protoplanet as does $M_b.$
Accordingly we shall consider the effects of asymmetric wave torques
as modifying the value of $M_b$ and continue to use (\ref{crto}).

Suppose now the protoplanet is acted on by some external torque $T_{ext}.$
The equation of motion governing the migration of the protoplanet
of mass $M_p$ with speed $dR/dt$ 
 obtained by considering the conservation of angular
momentum  is

\be {1\over 2} M_p R \Omega_p (dR/dt) = {1\over 2}( M_d - M_b)R\Omega_p 
(dR/dt) + T_{ext}.
\label{crto1}.\ee
Accordingly we can consider the planet to move with an effective mass

\be M_{eff} = M_p- ( M_d - M_b). \ee
The quantity $( M_d - M_b)$ has been called the coorbital mass deficit
by Masset \& Papaloizou (2003). When this is positive
there is an effective reduction in the inertia of the protoplanet.
If there were no asymmetry in wave torques, the coorbital mass deficit 
is seen to
be the amount of mass evacuated in the gap region were it to be initially
filled with the background density. Gap filling accordingly reduces
the coorbital mass deficit.
It is also clear that because at least a partial
gap is required, the protoplanet must be  massive enough to
produce a non linear response in the disk. Hence we have considered $q=2.$

Masset \& Papaloizou (2003) indeed found the  coorbital mass deficit 
to be positive resulting in positive feedback
from coorbital torques. They also found that in some circumstances
the coorbital mass deficit could become as large as the planet mass
so reducing the effective inertia to zero. Under these circumstances
a very fast or runaway migration may occur.

We have considered four simulations with torqued disk material
giving flow through the coorbital
region which correspond to a migrating protoplanet.
These are model G with $q=2.0, b/H =0.3$ and $\epsilon/h = 3/(8\pi),$
model H with $q=2.0, b/H =0.3$ and $\epsilon/h = 3/(64\pi),$
model I with $q=2.0, b/H =0.6$ and $\epsilon/h = 3/(8\pi),$ and
model J with $q=2.0, b/H =0.6$ and $\epsilon/h = 3/(64\pi).$
All of these models are inviscid with no magnetic field.
Results for models with flow through and 
magnetic fields will be presented elsewhere.
They all had $q=2.0$ which, for $h=0.05,$
 corresponds to a mass ratio of $2.5 \times 10^{-4}$
which is close to a Saturn mass for a central solar mass.
In each case the breakdown of the box symmetry 
resulted in a net torque acting on
the protoplanet.

Typical density contour plots
for model G  left panel and model H  right panel 
obtained after a quasi steady state had been attained
are given in figure \ref{fig01}.
In these cases the magnitude of the torque applied to the disk material
was chosen to produce  expected inflow speeds of
$(3/(8\pi))c_s$ and $(3/(64\pi))c_s$ respectively.
The effective introducing a flow through the coorbital region is
to tend to fill in the gap and produce an asymmetry in the density profile,
these effects being larger for a faster flow. In particular there is a density 
enhancement leading the planet that leads to a positive torque on the protoplanet
assisting its outward migration.

The mean surface density averaged over $y$ and $z$
as a function of radial
coordinate $x/H$ for $q=2.0$ with no flow through  (model F)
and  for model G are plotted in figure \ref{fig20}.
It is seen that the effect of the flow is to partially fill the gap
and to produce a density asymmetry that gives a relative density increase
just beyond the outer gap edge. This is responsible for an increase in the torque
coming from these regions that acts against the sense of protoplanet migration
and can be viewed as reducing the coorbital mass deficit.

Mean surface density averaged over $y$ and $z$
as a function of radial
coordinate $x/H$ for models G, H, I, and J are plotted in 
figure \ref{fig2}. These plots show an increasing amount of gap filling
as the flow increases. But note that this effect is less pronounced
for the smaller softening parameter because of the more effective gap production
in that case.

Theoretical considerations  (Masset \& Papaloizou 2003)
suggest the importance of trapped librating
coorbital material.
Velocity vectors in the coorbital region for model F
with no flow through are plotted 
in figure \ref{fig10}. This indicates trapped librating 
trajectories in the coorbital zone  essentially
 symmetrically placed with respect to the 
protoplanet and occupying a region of width $ w \sim 2H.$
For comparison velocity vectors in the coorbital region for   model G 
with fast flow through induced  by torqued  disk material.
are plotted in figure \ref{fig3a}. This shows a trapped region of librating
orbits predominantly leading the protoplanet
and shifted inwards.

Running time averages of the force acting on the planet in the direction
of orbital motion as a function of dimensionless time for
models G, H, I and J are
plotted in figure \ref{fig3}.
For comparison with theoretical expectations,
equation (\ref{crt00}) gives
in our arbitrary units for the original background
density and for the fast flow through rate 
with $(dR/dt)/c_s = 3/(8 \pi),$  $F_{cr} = 3.0 \times 10^{-11}.$   
 Simarly for the slow flow through rate with $(dR/dt)/c_s = 3/(64 \pi)$  we get
$F_{cr} = 4.0 \times 10^{-12}.$ 
We see that the latter  net running means are consistent with these values.
After allowing for adjustment of the  background value,
model G is a factor of two smaller, model H is a factor of two smaller, model
I is a factor of eight smaller while model J is a factor of six smaller.
We see therefore that, consistently with their more
pronounced gaps,  the cases with smaller softening give larger
coorbital torques. This appears to be because gap clearance is more effective 
leading to a larger coorbital mass deficit.
In addition the faster flow rates show indication of gap filling
and consequent tendency towards
 torque saturation. An interesting aspect of these
net torques is that in the faster flow through cases, the final net torque is 
comparable
to the initial one sided torque. This means that the resulting
 net torque is comparable to the extrapolated type I value assuming no gap (see also Masset \&
Papaloizou 2003).

\subsection{Occurrence of Fast or Runaway Migration}
Fast or runaway migration occurs when the effective mass becomes 
very small or zero (Masset \& Papaloizou 2003).
Then the migrating planet loses inertia and 
can migrate rapidly in response to even small external torques.
This requires $M_p= ( M_d - M_b).$  If we set $( M_d - M_b) =f M_d,$
$f$ is given above as $0.5$ for models G, and H ,$0.17$ for model J ,
and $0.125$ for model I respectively. 
The smaller cases apply for the larger softening parameter.
To obtain  $ M_{eff}=0.$ one requires $M_d =M_p/f.$ This means
the amount of background disk material required to fill the gap region
should be $M_p/f.$ One might expect this to be achievable for a massive
enough disk. For example taking the case with the weakest coorbital
torque with $f=0.125,$ $M_p= 2.5\times 10^{-4}M_{\odot}$ corresponding
to $q=2$ for $h =0.05,$ one requires a disk about $20$ times more massive
than the minimum mass solar nebula in the $5AU$ region. 
This is  essentially consistent with Masset \& Papaloizou (2003)
but cautionary remarks need to be inserted regarding the dependence
on softening and  the extrapolation of the behaviour in the local box to the
whole disk annulus.

\section { Discussion} \label{S5}
In this paper we have reviewed  local disk simulations
performed using a local shearing box. 
Such simulations 
have an advantage
of giving higher resolution for the same computational cost
as a global simulation, but with their associated boundary condition
of periodicity in shearing coordinates possibly introduce artefacts
due to neighbouring ghost boxes and protoplanets.
 However, we saw no evidence
that such features, although they can never
be removed entirely, were very significant.
 
We considered protoplanets interacting with
both quiescent disks and 
disks undergoing MHD turbulence with zero net flux fields.
The latter case results in angular momentum transport 
with an effective Shakura \& Sunyaev (1973)
parameter $\alpha$ related to the effective kinematic
viscosity $\nu$ by $\nu = \alpha H^2 \Omega_p,$
given by  $\alpha \sim 5\times 10^{-3}.$ 

In both cases, when a protoplanet is present, 
there exists a natural scaling indicating that results
depend only on the parameter  $q = M_p R^3/(M_* H^3)$
and the condition for gap formation is
$M_p R^3/(M_* H^3) > \sim 1 ,$
 being the  thermal condition  of
Lin \& Papaloizou (1993).

The symmetry of a shearing box for a non migrating protoplanet
 ensures cancellation of the torque contributions
from exterior and interior to the orbit. In a
quiescent disk a natural bias occurs when non local
 curvature effects are introduced. The interaction is then stronger
in the region exterior to the planet leading to a net inward migration
that can be estimated  using the expression of
Tanaka,Takeuchi,\& Ward (2002) for the migration time
\begin{equation}
\tau_{mig}= \frac{M_*}{2.7M_p}\frac{M_{*}}{{\Sigma R^2}}
h^2 \Omega_p^{-1} .\label{ward-tmigr}
\end{equation}
However, the torques acting in a turbulent disk from either
side of the orbit  are very noisy
to the extent that a ten percent difference between the contributions
can remain even after fifty orbits. The noise makes type I migration stochastic
and evaluation of the final outcome will require
long time global simulations which can simulate the fluctuations
that occur on the longest timescales associated with the disk
(see Nelson \& Papaloizou 2004).

We then considered
the coorbital torques experienced by a moving protoplanet
in an inviscid disk.
This was done by 
demonstrating, in an appropriate
slow migration limit,  the equivalence of the problem for a moving
protoplanet to one where the protoplanet is in a fixed orbit which the
disk material flows through radially as a result of the action of
an appropriate external torque.

By considering the torques exerted
by material flowing around and  through the coorbital region, 
we showed that we could regard the 
protoplanet as  moving  with an effective mass
\be M_{eff} = M_p- ( M_d - M_b). \ee
The quantity $( M_d - M_b)$ was  called the coorbital mass deficit
by Masset \& Papaloizou (2003). When this is positive
there is an effective reduction in the inertia of the protoplanet
a by this amount. It is related to, but not exactly equivalent to,
 the amount  
of mass evacuated in the gap region were it to be initially
filled with the background density. Because at least a partial
gap is required the protoplanet must be large enough to
produce a non linear response in the disk. 

In addition in order to
obtain measurable quasi steady 
coorbital torques a quasi steady state must be realized
in which the planet  does not accrete
significant mass. We found as did 
 Masset \& Papaloizou (2003) that
the coorbital torques are
proportional to the migration speed, when that is sufficiently
small,  and result in a positive
feedback on the migration, enhancing it and potentially leading to a runaway.
This could lead to fast migration for protoplanets in the Saturn mass
range in massive disks and may be relevant to the mass period correlation
for extrasolar planets which gives a preponderance of sub Jovian masses
at short orbital periods ( see Zucker \& Mazeh 2002).

The fast migration is limited in that the time to
migrate through the coorbital region of $w \sim 2H,$
should be less than the time it takes material to circulate around
the trapped coorbital region. When this is taken to extend
for the full $2\pi,$  this limit gives 
$R/(dR/dt) > 2\pi(R/H)^2/(3\Omega_p)$ which corresponds to a time
on the order of a hundred orbits. 
Thus a migration stopping or slowing  mechanism is required.
This may be as a result of the radial density profile of the disk
such as for example  a  reduction  of the density at smaller radii
 and /or a transition to type II
migration which operates on the longer viscous time scale
(see Masset \& Papaloizou 2003  for more discussion).


\section*{References}

\noi
Artymowicz, P., 1993,\apj, 419, 166

\noi
Balbus, S. A.,  Hawley, J. F., 1991,
ApJ, 376, 214

\noi
Boss, A.P., 2001, ApJ, 563, 367

\noi
Bryden, G., Chen, X., Lin, D.N.C., Nelson, R.P., Papaloizou, J.C.B, 1999, \apj
514, 344

\noi
D'Angelo, G.,  Henning, Th., Kley, W., 2002,  ApJ, 385, 647

\noi 
Goldreich, P.; Lynden-Bell, D., 1965, MNRAS, 130, 159

\noi
Goldreich, P., Tremaine, S.D., 1979, \apj, 233, 857

\noi
Hawley, J. F., Gammie, C. F.,  Balbus, S. A., 1995, ApJ, 440, 742

\noi
Hawley, J. F.,  Stone, J. M., 1995,
Computer Physics Communications, 89, 127

\noi
Kley, W., 1999, MNRAS, 303, 696

\noi
Korycansky, D. G.; Papaloizou, J. C. B., 1996, ApJS, 105, 181

\noi
Lin, D.N.C., Papaloizou, J.C.B., 1986, ApJ, 309, 8

\noi
Lin, D.N.C., Papaloizou, J.C.B., 1993,
Protostars and Planets III, p. 749-835

\noi
Marcy, G. W., Cochran, W. D., Mayor, M., 2000,
Protostars and Planets IV (Book - Tucson: University of Arizona Press;
eds Mannings, V., Boss, A.P., Russell, S. S.), p. 1285

\noi
Lubow, S.H., Seibert, M., Artymowicz, P., 1999, ApJ, 526, 1001

\noi
Masset, F.S., 2001, ApJ, 558, 453

\noi
Masset, F.S., Papaloizou, J.C.B., 2003, ApJ, 588, 494

\noi
Mayor, M., Queloz, D., 1995, Nature, 378, 355

\noi
Nelson,R.P., Papaloizou, J.C.B., Masset, F.S., Kley, W., 2000,
 MNRAS, 318, 18

\noi
Nelson, R.P., Papaloizou, J.C.B., 2003, MNRAS, 339, 993 

\noi
Nelson,R.P., Papaloizou, J.C.B., 2004, MNRAS, 350, 849

\noi
Papaloizou, J.C.B, Lin, D.N.C., 1984, ApJ, 285, 818

\noi
Papaloizou, J.C.B., Nelson, R.P., Snellgrove, M.D., 2004, MNRAS, 350, 829

\noi
Papaloizou, J.C.B., 2002, A\&A, 338, 615

\noi
Pollack, J.B., Hubickyj, O., Bodenheimer, P., Lissauer, J.J., Podolak, M.,
Greenzweig, Y., 1996, \icar 124, 62

\noi
Shakura, N. I., Sunyaev, R. A., 1973, A\&A, 24, 337

\noi
Steinacker, A., Papaloizou, J.C.B., 2002, ApJ, 571, 413

\noi
Tanaka, H., Takeuchi, T., Ward, W.R.,  2002, ApJ, 565, 1257

\noi
Terquem, C., Papaloizou, J.C.B., Nelson, R. P. 2000, SSRv, 92, 323

\noi
Vogt, S. S., Butler, R. P.,
Marcy, G. W.; Fischer, D. A.;
Pourbaix, D., Apps, K.,
Laughlin, G., 2002, ApJ, 568, 352

\noi
Ward, W., 1997, Icarus, 126, 261

\noi
Ward, W., 2000,
Protostars and Planets IV (Book - Tucson: University of Arizona Press;
eds Mannings, V., Boss, A.P., Russell, S. S.), p. 1485

\noi
Winters, W., Balbus, S., Hawley, J.,  2003a, ApJ, 589, 543

\noi
Winters, W., Balbus, S., Hawley, J.,  2003b, MNRAS, 340, 519

\noi
Ziegler, U.,  R\"udiger, G., 2000, A\&A, 356, 1141

\noi
Zucker, S. Mazeh, T., 2002, \apj, 568, L113




{}
\end{document}